\documentclass[10pt,conference]{IEEEtran}
\IEEEoverridecommandlockouts
\usepackage{cite}
\usepackage{amsmath,amssymb,amsfonts}
\usepackage{algorithmic}
\usepackage{graphicx}
\usepackage{textcomp}
\usepackage{xcolor}
\def\BibTeX{{\rm B\kern-.05em{\sc i\kern-.025em b}\kern-.08em
    T\kern-.1667em\lower.7ex\hbox{E}\kern-.125emX}}
    
\usepackage{xargs}                      
\usepackage[colorinlistoftodos,prependcaption,textsize=tiny]{todonotes}

\usepackage{bbm}

\usepackage{comment}
\usepackage{graphicx}
\usepackage{amsmath}
\usepackage{amssymb}
\usepackage{pifont}

\usepackage[american]{babel}
\usepackage{microtype}
\usepackage{subcaption}
\usepackage[ruled,,linesnumbered ]{algorithm2e}
\SetAlFnt{\small}
\SetAlCapFnt{\small}

\usepackage{booktabs}%
\usepackage{color}%
\usepackage{xcolor}%
\usepackage{hyperref}

\usepackage[flushleft]{threeparttable} 
\usepackage[normalem]{ulem}
\usepackage{tabularx}
\usepackage{multirow}
\usepackage{footmisc}
\usepackage{mathtools}

\newcount\DraftStatus  
\usepackage{color}
\definecolor{darkgreen}{rgb}{0,0.5,0}
\definecolor{purple}{rgb}{1,0,1}
\definecolor{todocolor}{rgb}{0.9,0.1,0.1}
\definecolor{hycolor}{rgb}{0.7,0.7,0.3}
\definecolor{fixcolor}{rgb}{0.1,0.7,0.3}
\definecolor{graybg}{rgb}{0.95,0.95,0.95}

\newcommand{\draftnote}[2]{\ifnum\DraftStatus=1
	\marginpar{
		\tiny\raggedright
		\hbadness=10000
		\def\baselinestretch{0.8}
		\textcolor{#1}{\textsf{\hspace{0pt}#2}}}
	\fi}

\newcommand{\nbc}[3]{
	{\colorbox{#3}{\bfseries\sffamily\scriptsize\textcolor{white}{#1}}}
	{\textcolor{#3}{\sf\small$\blacktriangleright$\emph{#2}$\blacktriangleleft$}}
}

\newcommand{\hy}[1]{\nbc{HY}{#1}{hycolor}}

\newcommand{\tool}{\textsc{XLIR}\xspace}

\begin{document}
\newcolumntype{L}[1]{>{\raggedright\arraybackslash}p{#1}}
\newcolumntype{C}[1]{>{\centering\arraybackslash}p{#1}}
\newcolumntype{R}[1]{>{\raggedleft\arraybackslash}p{#1}}

\title{Cross-Language Binary-Source Code Matching with Intermediate Representations} 

\author{\IEEEauthorblockN{Yi Gui$^1$, Yao Wan$^{1*}$\thanks{$^*$Yao Wan is the corresponding author.}, Hongyu Zhang$^2$, Huifang Huang$^3$, Yulei Sui$^4$, Guandong Xu$^4$, Zhiyuan Shao$^1$, Hai Jin$^1$}
	\IEEEauthorblockA{$^1$National Engineering Research Center for Big Data Technology and System, Services Computing Technology\\and System Lab, Cluster and Grid Computing Lab, School of Computer Science and Technology, \\Huazhong University of Science and Technology, Wuhan, China\\
	$^2$The University of Newcastle, Australia\\
	$^3$School of Mathematics and Statistics, Huazhong University of Science and Technology, Wuhan, China\\
	$^4$School of Computer Science, University of Technology Sydney, Australia\\
	\{guiyi, wanyao, hhf, zyshao, hjin\}@hust.edu.cn, hongyu.zhang@newcastle.edu.au, \{yulei.sui, guandong.xu\}@uts.edu.au}
}

\maketitle

\begin{abstract}
Binary-source code matching plays an important role in many security and software engineering related tasks such as malware detection, reverse engineering and vulnerability assessment. Currently, several approaches have been proposed for binary-source code matching by jointly learning the embeddings of binary code and source code in a common vector space. Despite much effort, existing approaches target on matching the binary code and source code written in a single programming language. However, in practice, software applications are often written in different programming languages to cater for different requirements and computing platforms. Matching binary and source code across programming languages introduces additional challenges when maintaining multi-language and multi-platform applications. To this end, this paper formulates the problem of cross-language binary-source code matching, and develops a new dataset for this new problem. We present a novel approach \tool, which is a Transformer-based neural network by learning the intermediate representations for both binary and source code. To validate the effectiveness of \tool, comprehensive experiments are conducted on two tasks of cross-language binary-source code matching, and cross-language source-source code matching, on top of our curated dataset. Experimental results and analysis show that our proposed \tool with intermediate representations significantly outperforms other state-of-the-art models in both of the two tasks.
\end{abstract}

\section{Introduction}
Binary-source code matching, which aims to measure the similarity between binary code and source code, plays an important role in a variety of security software engineering related tasks, e.g., malware detection~\cite{yarlagadda2020approach}, vulnerability search~\cite{yang2017bmxnet}, and reverse engineering~\cite{miyani2017binpro,shahkar2016matching}.
From one hand, given a binary code fragment, it is useful to retrieve similar source code snippets that can serve as references for reverse engineering.
On the other hand, given a vulnerable source code, it is also helpful to check whether its corresponding binary form is included in a binary file, which is useful for vulnerability assessment and detection.

\noindent\textbf{\textit{Existing Efforts and Limitations.}}
The core technique for binary-source code matching is the calculation of semantic similarity across two modalities (i.e., binary and source code).
To the best of our knowledge, most current methods are mainly concerning the matching within single modality, e.g., either source-to-source code matching~\cite{zhao2018deepsim}, or binary-to-binary code matching~\cite{yu2020order}.
Recently, several works have been proposed to investigate the binary-to-source matching problem. 
Yuan et al.~\cite{yuan2019b2sfinder} and Miyani et al~\cite{miyani2017binpro} studied the binary-to-source matching for open-source software reuse detection and binary source code provenance, respectively. 
Yu et al.~\cite{yu2020codecmr} studied the cross-modal matching of binary and source code at the function level. 
Both of these approaches extracted the semantic features of source code and binary code, and proposed two encoder networks to represent them as two hidden vectors. 
A similarity constraint (e.g., triplet loss function) is then designed to jointly learn these two encoder networks. 

Despite much progress having been achieved on binary-source code matching, all current works are exclusively developed to detect binary-source clones from programs written in the same programming language. However, detecting binary-source code clones for programs written in different programming languages has made little progress in literature. 
In practice, software applications are often written in different programming languages to cater for different platforms.
Therefore, detecting binary-source code clones across multiple programming languages is useful in real-world scenarios. 
For example, when we have a vulnerable binary code, it is necessary to retrieve all the relevant source code snippets for all possible programming languages they are written in, for better vulnerability assessment.
To fill this gap, we, for the first time, formulate the problem of \textit{cross-language binary-source code matching}.
\begin{figure}[!t]
	\centering
	\includegraphics[width=0.46\textwidth]{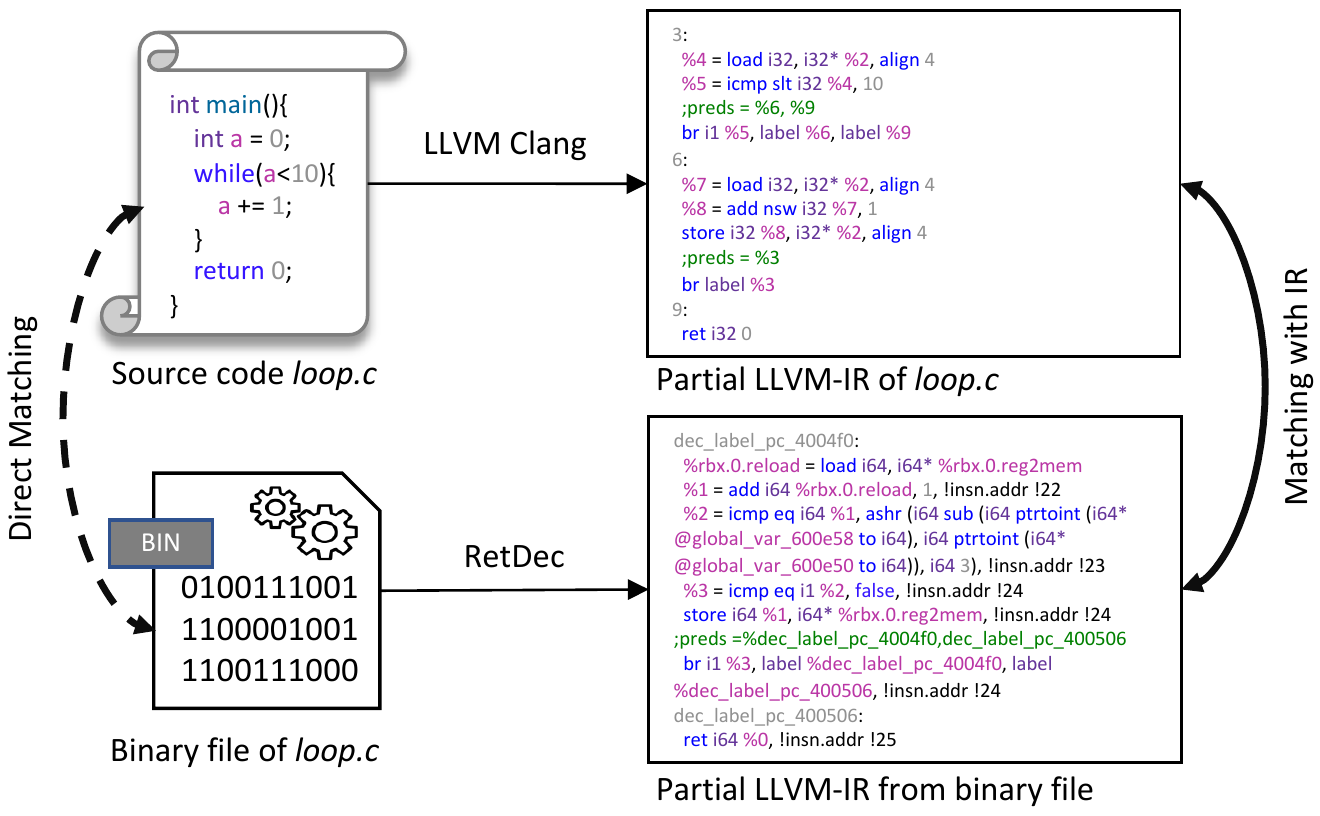} 
	\caption{The insights of representing binary code and source code using LLVM-IR. A simple loop is implemented in both the source code file and the binary file, and the LLVM-IRs generated from both of them indicate the similar semantics.
	}
	\label{fig_insights}
\end{figure}

\noindent\textbf{\textit{Insights.}}
The key challenge of binary-source code matching is to bridge the semantic gap between high-level programming language and low-level machine code even if they are in different textual appearance.
Current approaches aim to align the semantic embeddings of binary code and source code in an end-to-end way.
In compilers, intermediate representations are designed to support multiple front-end programming languages (e.g., C and Java) and multiple backend architectures (e.g., ARM and MIPS). 
That is, the intermediate representations are typically independent of programming languages and computer architectures, which can significantly reduce
the gap between binary code and source code by sharing a similar word vocabulary and the syntax structure.


For better illustration, Figure~\ref{fig_insights} shows  a source code fragment and a binary code fragment, together with their corresponding generated LLVM-IRs. We can transform the source code and binary code into LLVM-IR based on compiler tools, i.e., LLVM Clang\footnote{https://clang.llvm.org/\label{foot_clang}} and RetDec\footnote{https://retdec.com/\label{foot_retdec}}, respectively.
In this example, a \texttt{while} loop is implemented in both the source code file and the binary file.
It is hard to find their semantic similarity from the textual appearance. However, we can find some clues from their LLVM-IRs. Figure~\ref{fig_insights} (right) shows the partial LLVM-IRs generated from source code and binary code, both of which represent the semantics of the \texttt{while} loop fragment.
From other perspective, we can see that the intermediate representations can help unify the representations of source code and binary code of multiple programming languages and computer architectures.

\noindent\textbf{\textit{Our Solutions and Contributions.}}
Motivated by the aforementioned insights, this paper proposes \tool, a novel approach based on Transformer, for the task of cross-language binary-source matching using intermediate representations (IRs).
Specially, we parse both binary code and source code into intermediate representations.
In this paper, we adopt the intermediate representation i.e., LLVM-IR which has been widely used in compiler optimization, program analysis, bug detection and verification. The LLVM-IR can be translated from multiple high-level programming languages (e.g., C/C++ and Java) as well as low-level machine code.
To embed the intermediate representations, we adapt a Transformer-based neural network, which is first pre-trained by a masked language modeling on an external large-scale corpus of LLVM-IRs. We then map the LLVM-IR embeddings into a common space and jointly learn them using a triplet loss function.

To the best of our knowledge, there is no dataset for cross-language binary-source code matching. For evaluation, we curated and contributed a new comprehensive dataset based on an existing dataset originally developed for cross-language source-source code matching. Experimental results and analysis show that \tool significantly outperforms other state-of-the-art models. For the matching between Java binary code (compiled from corresponding LLVM-IR) and C source code, when comparing with the state-of-the-art tool B2SFinder, \tool significantly improves the Precision, Recall and F1 from 0.35, 0.41 and 0.38 to 0.68, 0.55 and 0.61, respectively.

Overall, this paper makes the following major contributions:
\begin{itemize}
    \item \textbf{New problem.} We, for the first time, formulate a new problem of cross-language binary-source code matching.
    \item \textbf{New insights.} Even though the source code and binary code are in different modalities, both of them can be transformed into IRs. In this paper, we provide an insight that it is feasible to mitigate the semantic gap between source code and binary code based on IRs (e.g., LLVM-IR).
    In addition to representing these LLVM-IRs, we propose an encoder network based on Transformer which is initialized by a pre-trained model for IR embedding.
    \item \textbf{Comprehensive experiments.} 
    To validate the effectiveness of our proposed approach, we first curate a new dataset based on a public dataset that is used for cross-language source code clone detection.
    We then conduct comprehensive experiments on two tasks of cross-language source-source code matching, and cross-language binary-source code matching, on top of our curated dataset.
    Experimental results validate the effectiveness of our proposed approach when comparing with the state-of-the-art models.
\end{itemize}

\begin{figure*}[!t]
	\centering
	\includegraphics[width=0.92\textwidth]{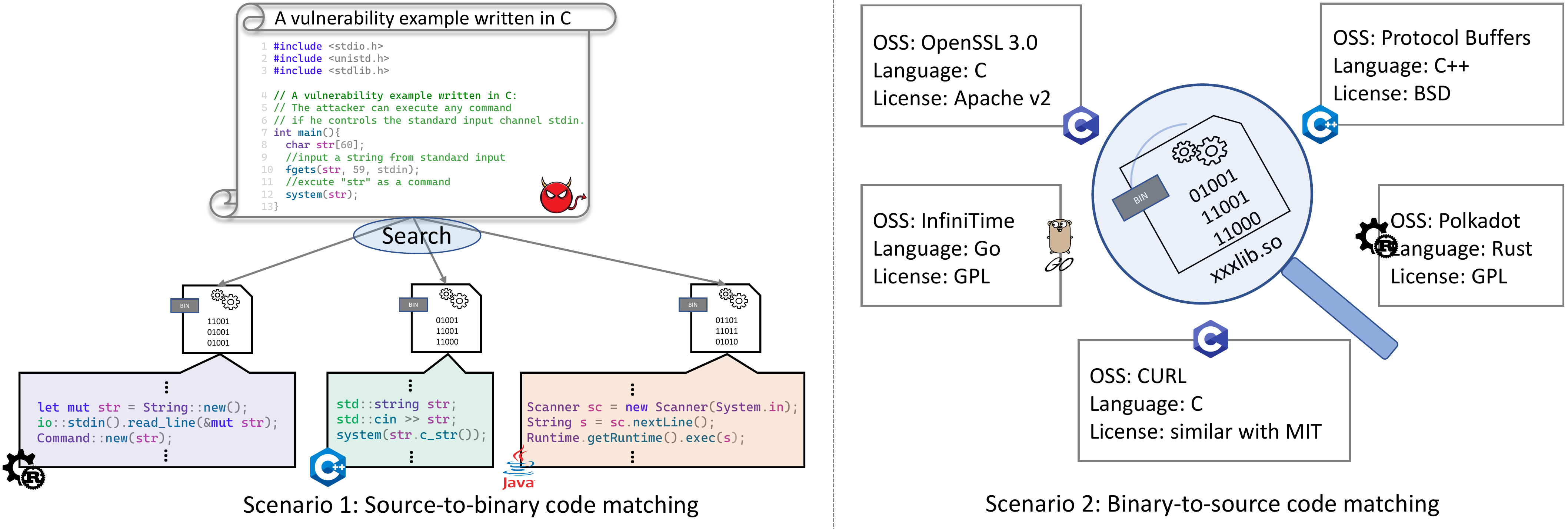} 
	\caption{The motivating scenarios of cross-language binary-source code matching. (1) A real-world scenario of source-to-binary code matching in vulnerability detection. In this case, given a source code with vulnerability, using which the attacker can execute any command if he/she controls the standard input channel \texttt{stdin}. 
	It is necessary for us to search for binary target files that may be written in other programming languages to check if there are similar backdoors in these binary target files. 
	(2) A real-world scenario of binary-to-source code matching in copyright protection. Given a binary library file, we match it to a known open source software (OSS) library (or some of its components) to determine whether the binary file is derived from it, and further check whether the binary file follows the license in the original OSS.
	}
	\label{fig_motivation}
\end{figure*}
\section{Motivation\label{sec_motivation}}
In this section, we first introduce the task of cross-language code clone detection, and then extend it to the cross-language binary-source code matching. We also present two practical scenarios of cross-language binary-source code matching.

\subsection{Cross-Language Code Clone Detection}
Detecting code clones is essential in software maintenance and refactoring. Existing efforts mainly focus on identifying code clones written in a single programming language.
With the emerging of multiple-language platforms where applications are often written in different programming languages to cater for different requirements and computing platforms.
It is useful to detect the code clones across different programming languages.
For example, in the collaborative development of software, when a code fragment is modified by a Java developer, it is required to propagate the changes to the corresponding code fragment written by a C developer.
Furthermore, when a developer finds a vulnerability in a code snippet written in one programming language, it is also useful to locate and identify the corresponding code fragment in other programming language. 
In cross-language code clone detection, the core insight is that although the source code fragments are written in different programming languages with distinct textual appearance, they may share similar semantics.

\subsection{Cross-Language Binary-Source Code Matching}
In this paper, we extend the cross-language code clone detection (i.e., cross-language source-source code matching) to the cross-language binary-source code matching.
We illustrate the motivation of this task using two real-world scenarios, as shown in Figure~\ref{fig_motivation}.

{\textit{Scenario 1: source-to-binary code matching in vulnerability detection.}} 
As shown in Figure~\ref{fig_motivation} (1), given a vulnerability source snippet written in C, in which the program accepts a string from the standard input stream and executes this string as a command. 
If the attacker has rights to write to the standard input stream, he/she can execute arbitrary commands on a target machine. 
The essence of this vulnerability is that the program directly uses the standard input stream as commands to execute, and this easy-to-implement pattern of backdoor may appear in existing binary files implemented by other programming languages. 
If we can match the vulnerable source code to binary code directly, we have a chance to find this vulnerability pattern in existing binary files.  

{\textit{Scenario 2: binary-to-source code matching in copyright protection.}} It is very common to introduce open source software (OSS) libraries written in different programming languages in software development, while different OSS libraries may use different open source licenses (e.g, GPL, Apache, BSD, MIT, etc.). Different licenses vary in details, for example, the GPL license requires that if a referrer modifies the original codes, it must also follow the GPL license. As shown in Figure~\ref{fig_motivation} (2), given a binary library file, if we can find which OSS (or components of it) it may be derived from, we can check whether it follows the protocol in the original library, which is very meaningful for copyright protection. In this scenario, cross-language binary-to-source code matching can play an important role.

Directly matching a binary file composed of machine instructions with a source code file requires professional knowledge and skills and is usually very complicated, so an end-to-end method to solve this problem is desired. 

\section{Preliminaries\label{sec_preliminaries}}

\subsection{Intermediate Representation (IR)\label{subsec_pre_ir}}
The \textit{Intermediate Representation} (IR) is a clearly-defined and well-formed  representation of programs with generally simple syntax rules, used by a compiler while making transformation from source to target.
Modern compilers first parse the source code, translate it into IR, and then generate target code from IR.
This additional layer has a bi-directional independent property, i.e., the IR is independent from both the source code and the target machine, while keeping the semantics of a program. Hence, the IR forms the basis of our cross-language matching method.
In this paper, we adopt the LLVM-IR, a specific type of IR first proposed by the LLVM infrastructure~\cite{lattner2004llvm}.
While containing the semantics of source code, the LLVM-IR is also in \textit{Static Single Assignment} (SSA) form~\cite{SSA}, in which any local variable is assigned exactly only once. 


Originally implemented for C and C++, the language-agnostic design of LLVM has since spawned a wide variety of front ends (e.g., C\#, and Rust).
Source code written in common programming languages can be easily compiled to LLVM-IR, and binary files can also be decompiled to LLVM-IR through some tools. We convert both the source code and binary files into LLVM-IRs, and then process them through an encoder to obtain the latent vectors. After calculating the similarity between the vectors, we can detect source-source, source-binary, or binary-source clones. In this way, we can achieve a universal end-to-end approach for code clone detection.





\subsection{Code Embedding}

Code embedding, also termed code representation learning, aims to preserve the semantics of programs into distributed vectors. It is essential in current deep learning-based program analysis.
To the best of our knowledge, current code embeddings fall into four categorizes according to the code features they represent: token sequence, AST, IR, and code graphs.
It is natural to represent the code based on its textual tokens, which reflect the lexical information of code.
To represent the structural information of code, several works~\cite{mou2014tbcnn,DBLP:conf/nips/ChenLS18,leclair2020improved,leclair2020improved,sui2020flow2vec,DBLP:conf/iclr/GuoRLFT0ZDSFTDC21} also propose to represent the AST and code graphs (such as control flow graph and data flow graph) using structural neural networks (e.g., TreeLSTM and GGNN).
Recently, several works~\cite{DBLP:conf/nips/Ben-NunJH18,venkatakeerthy2019ir2vec,brauckmann2020compiler} propose to represent the low-level information of code using the IRs. 

\begin{figure*}[!t]
	\centering
	\includegraphics[width=0.92\textwidth]{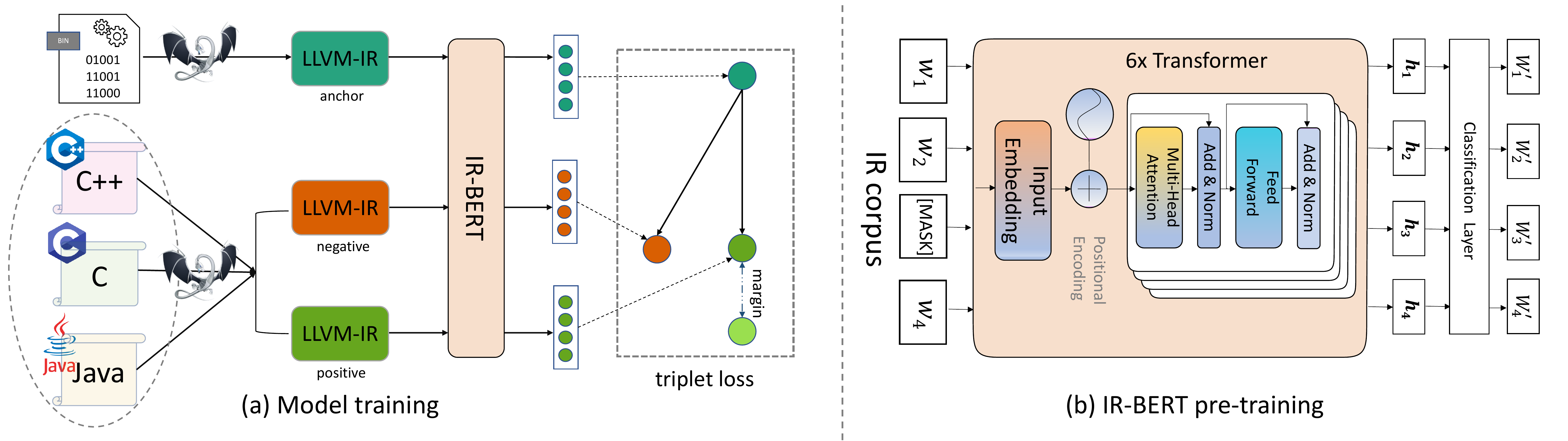} 
	\caption{An overview of our proposed \tool. In the model training phase (a), we first parse both the binary code and source code into IRs via several compiler tools. Then we feed the generated IRs into a pre-trained language model (i.e., IR-BERT based on Transformer for IR embedding (b). We learn the whole network by jointly mapping the embeddings of binary code and source code in a common space.}
	\label{fig:workflow}
\end{figure*}

\subsection{Problem Formulation}
Here we formulate the problem of cross-language binary-source code matching formally.
Note that, as an initial work, in this paper we limit our scope to matching the cross-language binary code and source code in files. Let $\mathcal{S}^{(l_u)} = \{s_1^{(l_u)}, s_2^{(l_u)},\ldots,s_n^{(l_u)}\}$ denote a set of source code files that are written in programming language $l_u$, $\mathcal{B}^{(l_v)} = \{b_1^{(l_v)}, b_2^{(l_v)},\ldots,b_n^{(l_v)}\}$ denote a set of binary code files that are built from the semantic-equivalent programs in $\mathcal{S}^{(l_u)}$ that are implemented in a different programming language $l_v$.
When $l_u$ and $l_v$ denote the same programming language, the studied task will be reduced into the binary-source matching task that has been studied in~\cite{yu2020codecmr}.
Given the paired source code files with their corresponding binary code files, the goal of this paper is to respectively learn the embeddings of both of them, and then align these embeddings in a common space.
The intuition is that the embeddings of paired source code and binary code in ground truth should be similar, while the embeddings of unpaired source code and binary code should be distinct as much as possible.


Formally, given a source code file $s_i^{(l_u)}$ and binary code file $b_j^{(l_v)}$, the embeddings of them are denoted as $\mathbf{s}_i^{(l_u)}$ and $\mathbf{b}_i^{(l_v)}$, respectively.
We map the embeddings of source code and binary into a common feature space via $\phi$ and $\Phi$, respectively.
\begin{equation}\label{eq_joint_embedding}
\small
    \mathcal{S}  \overset{\phi}{\rightarrow} V_{\mathcal{S}} \rightarrow J(V_{\mathcal{S}}, V_{\mathcal{B}}) \leftarrow V_{\mathcal{B}} \overset{\Phi}{\leftarrow} \mathcal{B}\,,
\end{equation}
where $J(\cdot,\cdot)$ denotes the similarity function, e.g., cosine similarity, which is designed to measure the matching degree of $V_{\mathcal{S}}$ and $V_{\mathcal{B}}$, in order to learn the mapping functions.

In this paper, we argue that source code and binary code are in distinct feature space, we propose to first map them into a closer feature space of IRs.
Generally, we can parse both source code and binary code into IRs, and then jointly learn their embeddings based on the IRs.
Therefore, the Eq.~\ref{eq_joint_embedding} can be reformulated as follows:
\begin{equation}\label{eq_joint_embedding_ir}
\small
    \mathcal{S} \overset{\text{parser}}{\longrightarrow} \mathcal{S}_{r} \overset{\phi}{\rightarrow} V_{\mathcal{S}} \rightarrow J(V_{\mathcal{S}}, V_{\mathcal{B}}) \leftarrow V_{\mathcal{B}} \overset{\Phi}{\leftarrow} \mathcal{B}_{r} \overset{\text{parser}}{\longleftarrow} \mathcal{B}\,.
\end{equation}
Eq.~\ref{eq_joint_embedding} and Eq.~\ref{eq_joint_embedding_ir} show that we can transform the problem of matching source code and binary code from their original textual representation to the mid-level IRs. 

\section{Cross-Language Binary-Source Code Matching\label{sec_steps}}
In this section, we present \tool, which is designed  for cross-language binary-source code matching with IRs.

\subsection{An Overview} 

Figure~\ref{fig:workflow} shows an overview of our proposed \tool. 
The model training phase is composed of the following three steps: (1) \textit{Transforming Source and Binary Code into IRs (cf. Sec.~\ref{subsec_ir}).} 
We first parse both the binary code and source code that are from different programming languages into IRs via several compiler tools (i.e., LLVM Clang\footnote{https://clang.llvm.org/} and JLang\footnote{https://polyglot-compiler.github.io/JLang/\label{foot_jlang}}).
Currently, we can support the C, C++ and Java programming languages.
(2) \textit{Transformer-based IR embedding (cf. Sec.~\ref{subsec_emb}).}
To represent the generated IRs, we feed them into a pre-trained Transformer-based language model (i.e., IR-BERT) for IR embedding.
We pre-train a masked language model on a large-scale IR corpus, following the CodeBERT~\cite{DBLP:conf/emnlp/FengGTDFGS0LJZ20} and OSCAR~\cite{DBLP:conf/icml/PengZLKHL21}, which are pre-trained models on code corpus and IR corpus, respectively.
(3) \textit{Model learning (cf. Sec.~\ref{subsec_model}).}
To correlate the embeddings of paired binary code and source code, we first map them into a common feature space, and jointly learn their correlations.

As the model trained, at the code matching phase, the cosine similarity is applied to measure the semantic similarity between binary code and source code. The matching score greater than a pre-defined threshold indicates that the binary code and source code from cross languages are matched.

\subsection{Transforming  Source and Binary Code into IRs\label{subsec_ir}}
Without loss of generality, we choose C, C++ and Java as source programming languages in cross-language scenario, since the parser has greatly developed in these mature procedural languages.
Additionally, we also choose the LLVM-IR as an intermediate representation, because
    (1) the LLVM-IR is source-independent, which yields different programming languages sharing the same semantics will keep similar IR structure;
    (2) the LLVM-IR is also target-independent and translation from LLVM-IR to any target-dependent assembly code is easy in practical;
    and 
    (3) the LLVM-IR is widely recognized by the community, which can ease the code transformation, e.g., decompilation, optimization, and semantics extraction.

For the sake of rigor, in order to avoid leakage of string information such as function and variable names into the binary file during compilation, we pass the ``\texttt{-s}'' parameter to compiler to strip out all debug information when compiling.

At the stage of training data preparation, we exploit a variety of tools to transform different program representations into LLVM-IR. The procedure applies to two major representations of program: source code and binary code.
For source code, we use LLVM Clang\footref{foot_clang} to emit LLVM-IR from C and C++, which is officially supported by LLVM community. We use Jlang\footref{foot_jlang} and Polyglot\footnote{https://www.cs.cornell.edu/projects/polyglot/} to translate Java into LLVM-IR.
For binary code, we use the RetDec decompiler\footref{foot_retdec} to convert non-obfuscated binary files into LLVM-IR. Obfuscation is the deliberate act of creating source or binary code that is difficult to understand, the semantic information in obfuscated code is considered harder to extract. However, there is only a limited range of code in this form. 

Note that, LLVM-IR has two different representations holding exactly the same semantic information. The \textit{bitcode} format is intentionally designed for machine processing, e.g., code transformation and automatic optimization. The \textit{machine IR} (MIR) format is human-readable and has been widely used in program debugging and analysis. For the sake of efficiency, we make use of the bitcode format as a universal representation of IR for code embedding.


\subsection{Transformer-based LLVM-IR Embedding\label{subsec_emb}}
We adapt Transformer~\cite{vaswani2017attention} for LLVM-IR embedding. Transformer is based on the self-attention mechanism and has become a popular model for source code embedding.
As shown in Figure~\ref{fig:workflow}(b), a Transformer model is composed of $K$ layers of blocks, which  can encode a sequence of instructions into contextual representation at different levels: $\mathbf{H}^{k}=[ \mathbf{h}_1^k, \mathbf{h}_2^k,\ldots,\mathbf{h}_{n}^k ]$, where $k$ denotes the $k$-th layer. 
For each layer, the layer representation $\mathbf{H}^{k}$ is computed by the $k$-th 
layer Transformer block $\mathbf{H}^{k} = \mathrm{ Transformer}_{k}(\mathbf{H}^{k-1})$, $k \in \{1,2,\ldots,K\}$.  

In each Transformer block, multiple self-attention heads are used to aggregate the output vectors of the previous layer. 
A general attention mechanism can be formulated as the weighted sum of the value vector $\mathbf{V}$ using the query vector $\mathbf{Q}$ and the key vector $\mathbf{K}$:
\begin{equation}
	\small
	\operatorname{Att}(\mathbf{Q}, \mathbf{K}, \mathbf{V})=\operatorname{softmax}\left(\frac{\mathbf{Q} \mathbf{K}^{T}}{\sqrt{d_{\text {model}}}}\right) \cdot \mathbf{V}\,,
\end{equation}
where $d_{\rm model}$ represents the dimension of each hidden representation. For self-attention, $\mathbf{Q}$, $\mathbf{K}$, and $\mathbf{V}$ are mappings of previous hidden representation by different linear functions, i.e., 
$\mathbf{Q} =\mathbf{H}^{l-1} \mathbf{W}_{Q}^{l}$, $\mathbf{K} =\mathbf{H}^{l-1} \mathbf{W}_{K}^{l}$, and  $\mathbf{V} =\mathbf{H}^{l-1} \mathbf{W}_{V}^{l}$, respectively. At last, the encoder produces a final contextual representation $\mathbf{H}^{L} = [\mathbf{h}^{L}_{1}, \ldots, \mathbf{h}^{L}_{n}]$, which is obtained from the last Transformer block.

\noindent\textbf{\textit{Pre-Training of IR-BERT.}} 
Following~\cite{DBLP:conf/icml/PengZLKHL21}, we first pre-train a masked language model on a large-scale external LLVM-IR corpus, termed IR-BERT, and then transfer the parameters into our model.
Given a LLVM-IR corpus, each LLVM-IR is first tokenized into a series of tokens, using Byte Pair Encoding (BPE~\cite{sennrich2015neural}).
Before pre-training, we first take the concatenation of two segments as the input, defined as $c_1=\{w_1, w_2,\ldots,w_n\}$ and $c_2=\{u_1, u_2, \ldots, u_m\}$, where $n$ and $m$ denote the lengths of two segments, respectively. 
The two segments are always connected by a special separator token \texttt{[SEP]}.
The first and last tokens of each sequence are always padded with a special classification token \texttt{[CLS]} and an ending token \texttt{[EOS]}, respectively.
The concatenated sentences are then fed into a Transformer encoder as input.
In the masked language modeling, the tokens of an input sentence are randomly sampled and replaced with the special token \texttt{[MASK]}.
In practice, we uniformly select 15\% of the input tokens for possible replacement. Among the selected tokens, 80\% are replaced with \texttt{[MASK]}, 10\% are unchanged, and the left 10\% are randomly replaced with the selected tokens from vocabulary~\cite{devlin2019bert}.
Without loss of generality, we adopt the pre-trained model in ~\cite{DBLP:conf/icml/PengZLKHL21}, which can be seen as a variant of IR-BERT.

\subsection{Model Learning\label{subsec_model}}


We learn \tool by mapping the embeddings of binary code and source code into a common space, with a similarity constraint.
The intuition is that if a binary code and a source code have similar semantics, their embeddings should be close to each other.
Let triplet $\langle b, s^+, s^- \rangle$ denote a training instance, in which for binary code $b$, $s^+$ denotes the corresponding source code in compilation (also termed positive sample or anchor), $s^-$ denotes a negative code snippet that is randomly chosen from the collection of all source code files.
When training on the set of $\langle b, s^+, s^-\rangle$ triplets, \tool predicts the cosine similarities of both $\langle b, s^+\rangle$ and $\langle b, s^-\rangle$ pairs and minimizes the ranking loss~\cite{gu2018deep} as follows: 
\begin{equation}
\small
	\mathcal{L}=\sum_{\langle{b}, {s}^{+}, {s}^{-}\rangle\in \mathcal{D}} \max (0, \alpha-{\mathrm{sim}} (\mathbf{b}, \mathbf{s}^{+})+{\mathrm{sim}} (\mathbf{b}, \mathbf{s}^{-}))\,,
\end{equation}
where $\mathcal{D}$ denotes the training dataset, $sim$ denotes the similarity score between the binary code and source code, and $\alpha$ is a small constant margin. 
$\mathbf{b}$, $\mathbf{s}^+$ and $\mathbf{s}^-$ are the embeddings of $b$, $s^+$ and $s^-$, respectively. 
In this paper, we adopt the cosin similarity function and set $\alpha$ to $0.06$ by default. 

\subsection{Code Matching}

At the inference phase, given a binary code $b$, as well a set of source code
files $\mathcal{S}$, 
For each source code file $s\in \mathcal{S}$, we first feed the binary code and source code files into our trained model and obtain their corresponding embeddings, denoted as $\mathbf{b}$ and $\mathbf{s}$. Then we calculate the matching score between $\mathbf{b}$ and $\mathbf{s}$ as follows:
%
\begin{equation}
\small
	\mathrm{sim}(b,s) = \cos (\mathbf{b}, \mathbf{s})=\frac{\mathbf{b}^{T} \mathbf{s}}{\|\mathbf{b}\|\|\mathbf{s}\|}\,,
\end{equation}
where $\mathbf{b}$ and $\mathbf{s}$ are the vectors of binary code and source code, respectively. 
If the matching score is larger than a threshold, we consider the pair of binary code and source code as matched, otherwise unmached.
Generally, 80\% is used as the similarity threshold for code clone detection. In our experiments, unless otherwise specified, this value is used as default. We also evaluate the impact of threshold in Section~\ref{sec_results}.

\section{Experimental Setup\label{sec_experiments}}

\subsection{Research Questions}
We conduct experiments to answer the following research questions.
\begin{itemize}
    \item \textbf{RQ1:}  Is our proposed \tool effective in cross-language binary-source code matching? 
    \item \textbf{RQ2:} What is the effectiveness of our proposed \tool in single-language binary-source code matching? 
    \item \textbf{RQ3:}  Can our approach  with IRs be extended to detect source-source code functionality clones effectively?
    \item \textbf{RQ4:}  What is the influence of major factors of \tool?
\end{itemize}


\subsection{Evaluated Tasks and Dataset}

\noindent\textbf{\textit{Cross-Language Source-Source Code Matching.}} 
This task aims to detect the source code clones across different programming languages, and has been studied previously in~\cite{nafi2019clcdsa}.
In~\cite{nafi2019clcdsa}, the authors introduced the CLCDSA dataset\footnote{https://github.com/Kawser-nerd/CLCDSA/}, which is composed of code snippets across four programming languages (i.e., C++, C\#, Java and Python), collected from two online judge and contest platforms (i.e., AtCoder\footnote{https://atcoder.jp/\label{foot:atcoder}} and Google CodeJam\footnote{https://codingcompetitions.withgoogle.com/codejam/\label{foot:codejam}}).
In this dataset, each programming problem is affiliated with multiple solutions implemented in different programming languages.
To validate the effectiveness our proposed \tool on cross-language source-source code matching, we choose C, C++, and Java as the studied languages.
Since our approach depends on IRs, all the source code files are supposed to be compiled, we filter out files that are unable to be successfully compiled, and divide the dataset into training, validation and test set according to a ratio of 6:2:2.
Table~\ref{tab:cross_data} shows the statistics of filtered dataset used in this paper.

\noindent\textbf{\textit{Cross-Language Binary-Source Code Matching.}} 
Currently, there is no available dataset for the evaluation of cross-language binary-source code matching.
To mitigate this gap, we curate a new dataset based on the CLCDSA dataset, by compiling the source code in one programming language into binary code, while keeping the source code in another programming language unchanged.
In particular, we compile each source file into binary files using different compilers (i.e., GCC and LLVM Clang), with multiple optimization options (i.e., \texttt{-O0}, \texttt{-O1}, \texttt{-O2}, and \texttt{-O3}), across multiple platforms (i.e., \texttt{x86-32},  \texttt{x86-64}, \texttt{arm-32}, and \texttt{arm-64}).
Consequently, each source code file is compiled into 32 different object files. 
Under the single-language setting, 
we conduct experiments mainly on POJ-104~\cite{mou2016convolutional}, which is a public dataset composed of about 50,000 programs written in C and C++, we also collected data from a variety of online judge and contest platforms.
In our dataset, each problem is affiliated with around 500 solutions.

\begin{table}[t!]
    \centering
    \caption{An overview of the CLCDSA~\cite{nafi2019clcdsa} dataset for cross-language binary-source code matching.}
    \begin{tabular}{ll|ccc} 
    \hline
                             &      & Train & Validation & Test  \\ 
    \hline
    \multirow{3}{*}{AtCoder} & C    & 4,772  & 1,672       & 1,743  \\
                             & C++  & 4,606  & 1,605       & 1,692  \\
                             & Java & 4,856  & 1,706       & 1,770  \\ 
    \hline
    \multirow{3}{*}{CodeJam} & C    & 1,168  & 409        & 447   \\
                             & C++  & 1,176  & 410        & 445   \\
                             & Java & 1,105  & 379        & 397   \\ 
    \hline
    Total                    &      & 17,683 & 6,181       & 6,494  \\
    \hline
    \end{tabular}   
    \label{tab:cross_data}
\end{table}

\noindent\textbf{\textit{Dataset for Pre-Training.}} 
We first pre-train an IR-BERT on a 
separate large-scale IR corpus.
We adopt the dataset that has been used in~\cite{DBLP:conf/icml/PengZLKHL21}, which is composed of eleven real world popular softwares (i.e., Linux-vmlinux, Linux-modules, GCC, MPlayer, OpenBLAS, PostgreSQL, Apache, Blender, ImageMagcick, Tensorﬂow, Firefox) from GitHub.
We can compile these softwares into LLVM-IRs using LLVM Clang with \texttt{-O0} optimization level.
Finally, 48,023,781 LLVM-IR instructions from 855,792 functions are obtained.

\begin{table*}[t!]
\centering
\caption{Performance of cross-language binary-source code matching (BinPro and B2SFinder do not support Java, and Java binary code refers to the binary code compiled from the corresponding LLVM-IR).}
\begin{tabular}{l|ccc|ccc}
\hline
          & \multicolumn{3}{l|}{C/C++ binary code with Java source code} & \multicolumn{3}{l}{Java binary code with C/C++ source code}  \\
          \cline{2-7}
          & Precision & Recall & F1  & Precision & Recall & F1                                      \\ 
\hline
BinPro & -         & -      & -                            & 0.36      & 0.37   & 0.36                                           \\
B2SFinder    & -         & -      & -                            & 0.35      & 0.41   & 0.38                                           \\
XLIR (LSTM)      & 0.62      & 0.53   & 0.57                         & 0.55      & 0.51   & 0.53                                           \\
XLIR (Transformer)      & 0.73      & 0.59   & 0.65                         & 0.68      & 0.55   & 0.61                                           \\
\hline
\end{tabular}
\label{tab:cross_source_bin}
\end{table*}
\begin{table}[t!]
\centering
\caption{Performance of single-language C++ binary code to C++ source code matching on POJ-104 dataset}
\begin{tabular}{l|ccc} 
\hline
          & Precision & Recall & F1  \\ 
\hline
BinPro & 0.38      & 0.42   & 0.40        \\
B2SFinder    & 0.43      & 0.46   & 0.44       \\
XLIR (LSTM)      & 0.67      & 0.72   & 0.69       \\
XLIR (Transformer)      & 0.85      & 0.86   & 0.85       \\
\hline
\end{tabular}
\label{tab:single-code-binary-cmp}
\end{table}
\begin{table*}[t!]
\centering
\caption{Performance of cross-language code clone detection.}
\begin{tabular}{l|ccc|ccc|ccc} 
\hline
        & \multicolumn{3}{c|}{LICCA} & \multicolumn{3}{c|}{\tool(LSTM)} & \multicolumn{3}{c}{\tool(Transformer)}  \\ 
\cline{2-10}
        & Precision & Recall & F1    & Precision & Recall & F1     & Precision & Recall & F1            \\ 
\hline
C\&C++    & 0.43      & 0.37   & 0.40  & 0.78      & 0.65   & 0.71   & 0.92      & 0.86   & 0.89          \\
C\&Java   & 0.31      & 0.29   & 0.30  & 0.62      & 0.51   & 0.56   & 0.75      & 0.55   & 0.63          \\
C++\&Java & 0.33      & 0.29   & 0.31  & 0.65      & 0.53   & 0.58   & 0.77      & 0.57   & 0.66          \\
\hline
\end{tabular}
\label{tab:cross_source_detection}
\end{table*}

\subsection{Baselines}
We evaluate our proposed \tool on two code matching tasks (i.e., cross-language binary-source code matching and cross-language source-source code matching).
For each task, the performance of our model is compared \tool with the following state-of-the-art baselines.

\noindent\textbf{\textit{Cross-Language Binary-Source Code Matching.}} 
We extend the baselines for binary-source code matching, from single-language setting to cross-language setting.
\begin{itemize}
    \item \textbf{BinPro}~\cite{miyani2017binpro} extracts function call graphs (FCGs) for both binary and source code, and uses a bipartite matching algorithm (i.e., Hungarian algorithm~\cite{li2016vulpecker}) to match them.
    \item \textbf{B2SFinder}~\cite{yuan2019b2sfinder} extracts seven features from three perspectives (i.e., strings, integers and control-flow) for both binary and source code, and introduces a weighted matching algorithm to match them.
    \item \textbf{\tool(LSTM)} is a variant of our proposed approach, in which  LSTM~\cite{greff2016lstm} is used to encode the IRs.
\end{itemize}

\noindent\textbf{\textit{Cross-Language Source-Source Code Matching.}}
We first extend our proposed \tool to the task of cross-language source-source code matching, and then compare \tool with the following baselines.
\begin{itemize}
    \item \textbf{LICCA}~\cite{vislavski2018licca} is a tool for detecting source-to-source code clones across programming languages based on syntactic and semantic features of code. It has been verified in Java, C, JavaScript, Modula-2 and Scheme.
    \item \textbf{\tool(LSTM)} is a variant of our proposed approach, in which the LSTM~\cite{greff2016lstm} is used to encode the IRs. This approach has been adapted to the task of cross-language source-source code matching.
\end{itemize}

\subsection{Evaluation Metrics}
We adopt recall, precision and F1-score for model evaluations, which have been widely used in text matching and information retrieval~\cite{powers2020evaluation}.
For a query code snippet (source or binary code), we call the corresponding cloned code snippet as positive, and the non-cloned code snippet as negative. 
Let $T_p$ and $F_p$ be the number of positive clones detected true and false,
and $F_n$ be the number of negative clones detected false, the precision $P$ and recall $R$ is calculated as follows:

\begin{equation}
\small
    P = \frac{T_{p}}{T_{p} + F_{p}},\,\,R =  \frac{T_{p} }{T_{p}+F_{n}}\,.
\end{equation}

We also use F1 for evaluation, which is defined as the harmonic mean of Precision and Recall. It can be interpreted as a trade-off between them. 
Formally, F1 is defined as follows:
\begin{equation}
\small
    F1 = 2\cdot \frac{P\cdot R}{P + R}\,.
\end{equation}

\subsection{Implementation Details}
We implement \tool with PyTorch 1.9. As mentioned above, the encoder for LLVM-IR is based on the Transformer. For Transformer, the settings of each encoder are the same as BERT~\cite{devlin2019bert}. For the masking strategy, we take random 15\% instructions from IR. For these selected instructions, we replace them with the \texttt{[MASK]} token and random characters with 80\% and 10\% probability, and keep them unchanged with 10\% probability. We set the hidden size and word embedding size to 256.
In pre-training phase, we generate a dictionary from the LLVM-IRs compiled from a large code corpus. When fine-tuning our model on the clone detection task, unrecognized characters will be replaced with $\langle$\texttt{UNK}$\rangle$.

We conducted experiments on a Linux server having four Tesla V100 GPU of 32GB memory.
Our model training procedure is distributed on all four GPUs, and the parameters are updated via ADM optimizer, with the learning rate of $1e\text{-}3$ for training. 
To prevent over-fitting, we use a dropout of 0.4.

\section{Experimental Results and Analysis\label{sec_results}}

\subsection{RQ1: Effectiveness of IR for Cross-Language Binary-Source Code Matching}
In order to verify the performance of our proposed model, we conduct experiments on our curated dataset based on CLCDSA. 
Table~\ref{tab:cross_source_bin} shows the performance of binary-source code matching between binary files and source files written in C/C++ 
and Java. 
From this table, we can see that our algorithm is capable of detecting cross-language binary-source code clones. 
In the task of C/C++ binary to Java source code matching, precision, recall and F1 are as high as 0.73, 0.59 and 0.65, respectively, and in the task of matching Java binary and C source code, they are 0.68, 0.55 and 0.61, respectively. 
We have noticed that C/C++ and Java, source code files and binary files are quite different. 
Because of these differences, our results are quite encouraging. The results reveal that although source code and binary file are written in different programming languages and in different forms, functionality clone can be detected after they are converted into LLVM-IR by our approach. We think this may be attributed to the equivalent conversion of semantic information when source code files and binary files are transformed into LLVM-IR.

\subsection{RQ2: Effectiveness of IR for Single-Language Binary-Source Code Matching}
Since the CLCDSA dataset with a total of only about 30,000 source files after screening is relatively small, we extended the experiment of binary-source clone detection to the single-language dataset POJ-104 to further verify the performance of our approach in binary-source code matching.  
Table~\ref{tab:single-code-binary-cmp} shows the performance of C++ binary code to C++ source code matching. We can see that our \tool outperforms all baselines by large margins, e.g., F1 score of XLIR is higher than BinPro and B2SFinder by 0.42 and 0.41, respectively. This shows the effectiveness of our method in binary-source code matching.

\subsection{RQ3: Extended Evaluation on Cross-Language Source-Source Code Matching}
We conduct code clone detection between C, C++ and Java, and the results are shown in 
Table~\ref{tab:cross_source_detection}. When the similarity threshold is 80\%, no matter our \tool uses Transformer or LSTM as the encoder, the performance exceeds LICCA by a large margin. 
Specially, \tool achieves an average precision of 0.81, an average recall rate of 0.77, and an average F1 score of 0.73. 
This shows that our method can effectively detect cross-language functional clones. We can also see that the similarity between C and C++ is higher than that of between C/C++ and Java, because the basic syntax of C and C++ is very similar.

\subsection{RQ4: Influence of Major Factors}

We study the influence of the following major factors on our approach separately, i.e., pre-training, Transformer encoder, compilation options, and similarity threshold.

\noindent\textbf{\textit{Impact of Pre-Training.}}
To verify the impact of pre-training, we conduct the experiment of cross-language code clone detection without pre-training, and the results are shown in Table~\ref{tab_cross_source_detectionxx}. 
Compared with the performance with pre-training in Table~\ref{tab:cross_source_detection},  we can see the performance of our approach without pre-training deteriorates by a small but perceptible margin, i.e., Precision, Recall, F1 decreased by a an average of 0.04, 0.03 and 0.03, respectively. 
This shows the obvious effectiveness of pre-training on external large-scale datasets.
\begin{table}[t!]
    \centering
    \caption{Performance of \tool on cross-language code clone detection without pre-training.}
    \begin{tabular}{l|ccc} 
    \hline
            & Precision & Recall & F1  \\ 
    \hline
    C\&C++    & 0.88      & 0.84   & 0.86       \\
    C\&Java   & 0.71      & 0.52   & 0.60        \\
    C++\&Java & 0.72      & 0.54   & 0.62       \\
    \hline
    \end{tabular}
    \label{tab_cross_source_detectionxx}
\end{table}

\noindent\textbf{\textit{Contribution of the Transformer Encoder.}}
To verify the effectiveness of Transformer encoder in our approach, we conduct experiments of cross-language binary-source clone detection and source-source clone detection using LSTM as encoder for LLVM-IR embedding instead of Transformer. The experimental results are shown in Tables~\ref{tab:cross_source_bin},~\ref{tab:single-code-binary-cmp},  and~\ref{tab:cross_source_detection}.
The three tables show that our model with Transformer encoder outperforms that with LSTM encoder in terms of all evaluation metrics, on all the three tasks.
We attribute it to the fact that LLVM-IRs are always long sequences, while the Transformer works better on representing them. 

\noindent\textbf{\textit{Influence of Compilation Options.}}
\begin{table}[t!]
\centering
\caption{Performance of code-binary clone detection on POJ-104.}
\setlength{\tabcolsep}{3pt} 
\begin{tabular}{c|c|ccc|ccc} 
\hline
\multicolumn{2}{c|}{\multirow{2}{*}{}} & \multicolumn{3}{c|}{X86}                                                 & \multicolumn{3}{c}{ARM}                                                  \\ 
\cline{3-8}
\multicolumn{2}{c|}{}                  & \multicolumn{1}{c}{Precision} & \multicolumn{1}{c}{Recall} & F1 & \multicolumn{1}{c}{Precision} & \multicolumn{1}{c}{Recall} & F1  \\ 
\hline
\multirow{4}{*}{gcc}   & \texttt{-O0}            & 0.87                           & 0.84                        & 0.85      & 0.84                           & 0.87                        & 0.85       \\ 
                       & \texttt{-O1}             & 0.84                           & 0.86                        & 0.85      & 0.86                           & 0.88                        & 0.87       \\ 
                       & \texttt{-O2}             & 0.89                           & 0.85                        & 0.87      & 0.83                           & 0.84                        & 0.87       \\ 
                       & \texttt{-O3}             & 0.89                           & 0.88                        & 0.88      & 0.88                           & 0.82                        & 0.85       \\ 
\hline
\multirow{4}{*}{clang} & \texttt{-O0}             & 0.85                           & 0.86                        & 0.85      & 0.83                           & 0.87                        & 0.85       \\ 
                       & \texttt{-O1}             & 0.83                           & 0.85                        & 0.84      & 0.81                           & 0.86                        & 0.83       \\ 
                       & \texttt{-O2}             & 0.84                           & 0.83                        & 0.83      & 0.84                           & 0.87                        & 0.85       \\ 
                       & \texttt{-O3}             & 0.87                           & 0.82                        & 0.84      & 0.89                           & 0.82                        & 0.85       \\
\hline
\end{tabular}
\label{tab:complilation}
\end{table}
\begin{figure}[t!]
    \centering
	\includegraphics[width=0.5\textwidth]{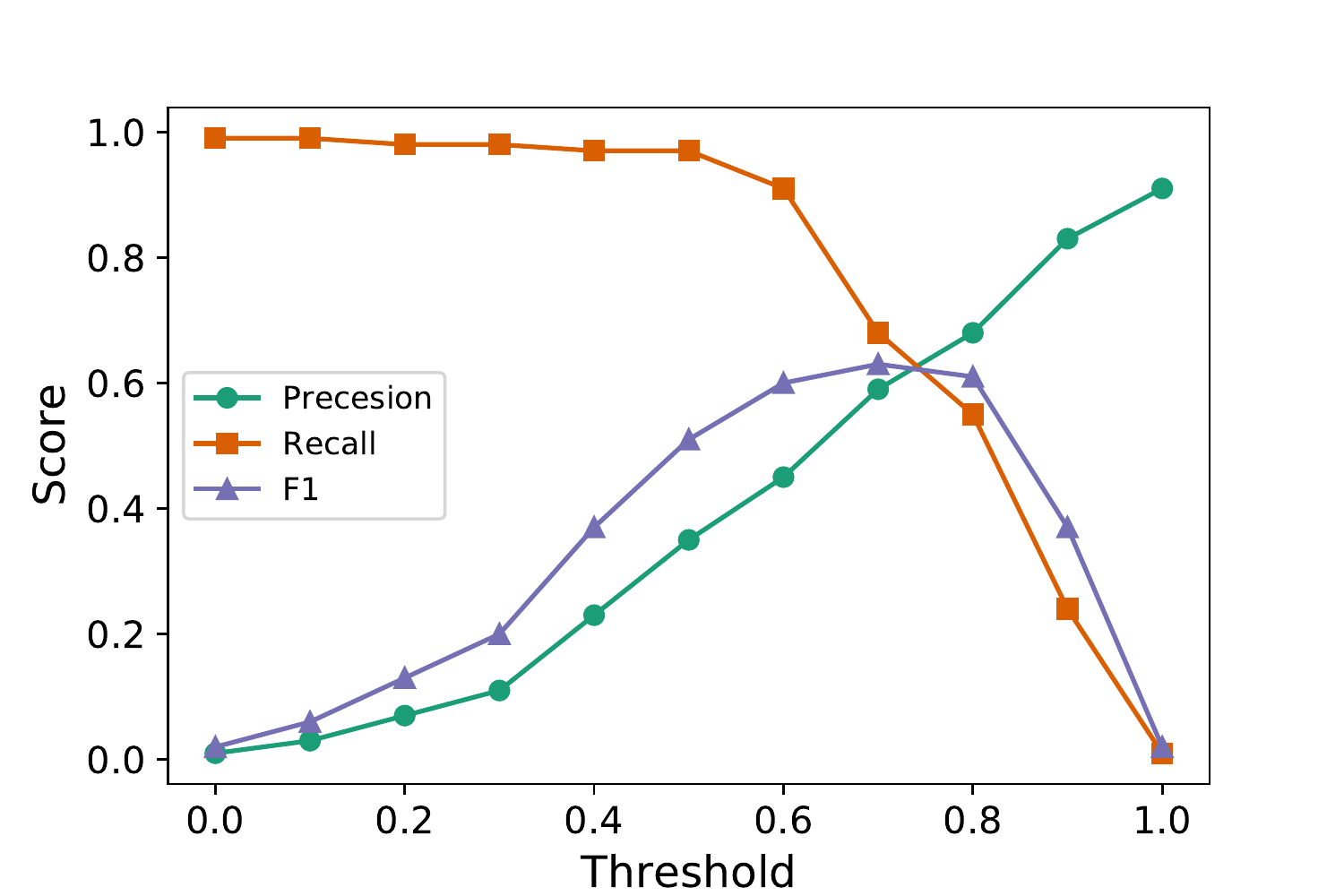}
	\caption{Influence of threshold on Java binary to C++ source code matching.
	}
	\label{fig:threshold}
\end{figure}
In order to investigate the impact of different compilation options on binary-source clone detection, we carried out an experiment on dataset POJ-104: we use different compilation options of compilers, platforms and optimisations to compile POJ-104 into  binary files (Linux elf files) , and then conduct clone detection between the source files and binary files. The results are shown in Table~\ref{tab:complilation}. We can see that our approach can consistently achieve good performance for various binary files. In particular, the average of precision, recall and F1 is about 0.85, and the variances are 0.0006, 0.0004, 0.0002, respectively. 
These results reveal that our approach is effective in matching source code and different binary files. 

\noindent\textbf{\textit{Influence of Threshold.}}
In our experiment, we set 0.8 as the default threshold value, that is, when the similarity is greater than or equal to 0.8, the code pair is considered to be cloned. 
In order to explore the impact of threshold values on the final results, we adjust the threshold and carry out a series of experiments on Java binary to C++ source code matching. The experimental results are shown in Figure~\ref{fig:threshold}.
From this figure, we can see that when the threshold rises from 0.5 to 0.98, the precision increases significantly while the recall drops quickly. 
This result is consistent with our intuition: the higher the similarity, the higher the precision; the lower the similarity, the higher the recall rate.
When the threshold is between 0.7 and 0.8, a balanced point is achieved. In our work, we set the threshold as 0.8 by default.

\section{Discussion \label{sec_discussion}}
\subsection{Case Study}
\begin{figure}[!t]
	\centering
	\includegraphics[width=0.5\textwidth]{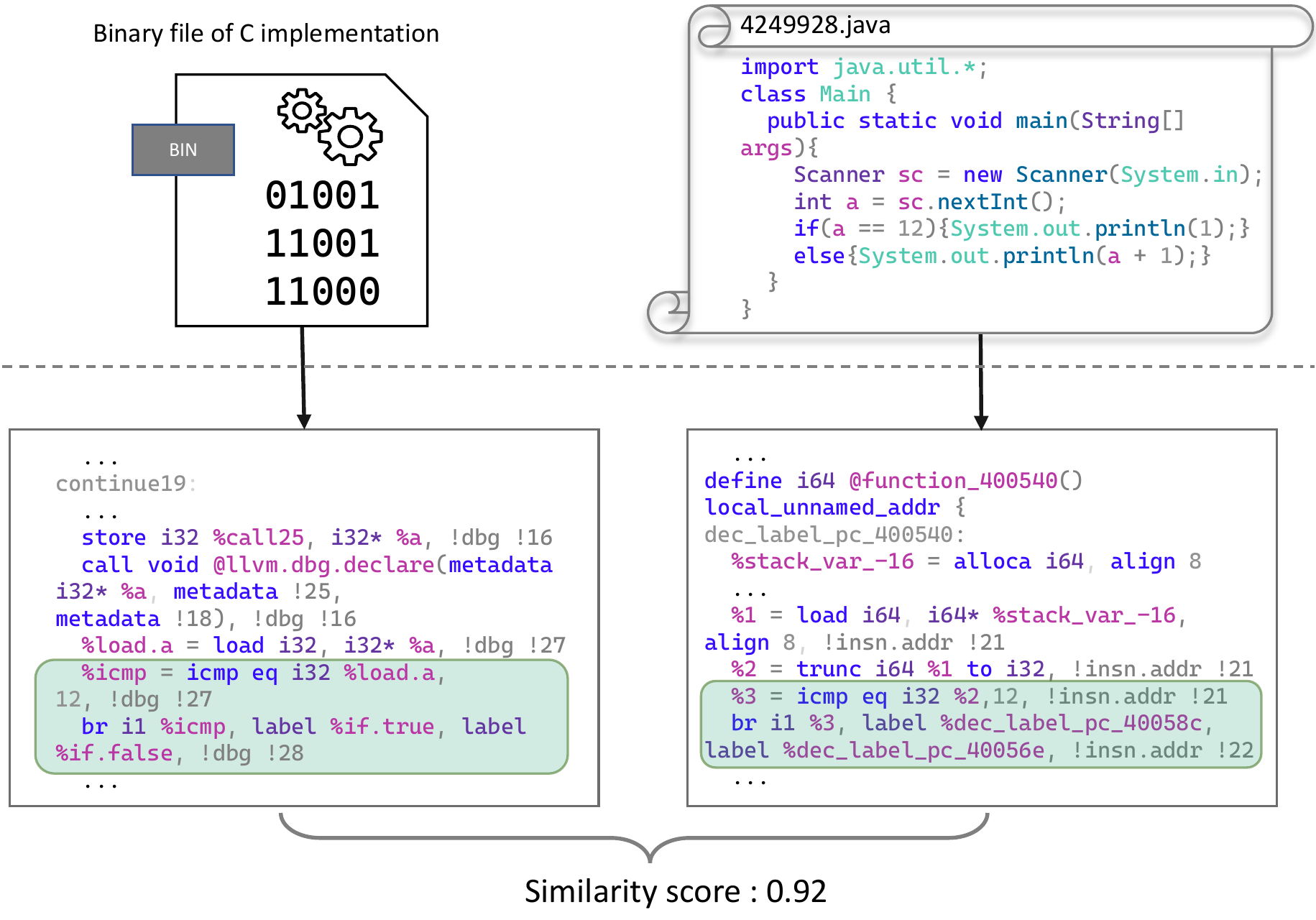} 
	\caption{A case study to show how Java source code matchs with C binary code. 
	}
	\label{fig_case_study}
\end{figure}
Figure~\ref{fig_case_study} shows a real example of how a C binary file matches a Java source file. The two images in the upper side of the figure show a C binary file and a Java source file for the same problem \textit{ABC011/A}\footnote{http://atcoder.jp/contests/abc011/tasks/abc011\_1/} in the programming contest website AtCoder. The problem is: 
\textit{given the input value $n$, if $n$ equals to 12, outputs $n+1$, otherwise outputs 1}. The two code snippets in the bottom of Figure~\ref{fig_case_study} are extracted from the LLVM-IR obtained by disassembling the binary object file and compiling the LLVM-IR of Java source code, respectively. We can see that the two LLVM-IR code snippets have equivalent semantics.
Both of them contain a branch, conditioned on whether the input value $n$ equals to 12 or not.
The previous input instructions and the subsequent output-related instructions are omitted based on the space limitation.
From this example we can see that an important reason why our method can work is that the semantic information is equivalently retained after converting to LLVM-IR.




\subsection{Strength of \tool}
We have identified two advantages of \tool: (1) Compared with some previous methods based on mainly extracting code literals in binary target file and specific characteristics in source code file, we transform source code file and binary target file into semantically equivalent LLVM-IRs for code matching task, which can greatly reduce the loss of program semantic information compared with extracting some manually selected features. (2) Our method is an end-to-end binary-source and source-source matching approach, so users do not need to grasp complicated reverse engineering skills to use this tool. This can greatly improve the efficiency of code clone detection and can be useful for software security related tasks.

\subsection{Threats to Validity and Limitations}\label{sec_threats}
The first threat is that the source code for clone detection in our approach must be compilable, 
and the binary object file can be disassembled into LLVM-IR. In a few scenarios, there may be cases where incomplete or grammatically incorrect code snippets cannot be compiled but still need to be checked. 
Furthermore, parsing a binary object file 
that is seriously obfuscated and cannot be decompiled has always been a challenging problem in reverse engineering and is not the focus of this paper. The tool RetDec we use is capable of decompiling binary object files to LLVM-IR in most cases.

Another limitation of our method is that it only supports programming languages that have static LLVM compilers. Although many programming languages can be easily compiled by LLVM compilers, such as C/C++, Rust, Java, Ruby, CUDA, LUA, Objective-C, C\#, OpenCL, etc., there are still some commonly-used programming languages that can not. For example, Python has only JIT's LLVM compiler due to the characteristic of being a dynamic programming language, that is, Python can only be translated into LLVM-IR during runtime. However, we believe that more and more programming languages will support static compilation to LLVM-IR, because LLVM has great potential in compiler optimization.



\section{Related Work\label{sec_related_work}}
\subsection{Deep Learning for Source Code}
The last few years have witnessed increasing interests on applying deep learning for source code modeling, so as to build intelligent tools to increase the productivity of software developers.
One fundamental task is code embedding, which can support a variety of downstream tasks, including code search~\cite{gu2016deep,wan2019multi-modal}, code summarization~\cite{hu2018deep,wan2018improving,DBLP:conf/iclr/AlonBLY19}, code completion~\cite{liu2016neural,svyatkovskiy2019pythia,kim2021code,liu2020self} and code clone detection~\cite{zhang2019novel,zhao2018deepsim,wu2020scdetector,hua2020fcca}.
From our investigation, current approaches mainly represent source code in four perspectives, i.e., sequential code tokens, ASTs, code graphs, and IRs. 
It is natural to simply represent the semantics of program as a sequence of tokens, like that in NLP. 
Current approaches mainly tokenize the program into sequential tokens by several special separators, e.g., whitespace or Camel cases (for identifiers like \texttt{SortList} and \texttt{intArray}).
To represent the structured syntactic information inside a program, one line of work use ASTs in the form of tree or graph for network feeding, this type of networks including Tree-CNN~\cite{mou2014tbcnn}, Tree-LSTM~\cite{DBLP:conf/nips/ChenLS18} and GGNN~\cite{leclair2020improved}.
Furthermore, another line of work represent the AST by serializing the structured AST into a list of instructions, so that sequential learning methods can be used~\cite{hu2018deep,alon2019code2vec}.
To add information from different perspective, a program can be translated into different representations. Augmented ASTs hold more detailed properties for a node~\cite{leclair2020improved}, data-flow graphs~\cite{DBLP:conf/iclr/GuoRLFT0ZDSFTDC21} and control-flow graphs~\cite{sui2020flow2vec} tend to skeletonize the program by highlighting execution paths and changing of variables.
Recently, some works resort to represent a program by IR, which is independent to programming languages and platforms~\cite{DBLP:conf/nips/Ben-NunJH18,venkatakeerthy2019ir2vec,brauckmann2020compiler}.
Benefited from the pre-training techniques in processing natural language tasks ~\cite{devlin2018bert}, Feng et al.~\cite{DBLP:conf/emnlp/FengGTDFGS0LJZ20} pre-trained CodeBERT for the bimodal of programming language and natural language, which has shown promising results in various code-related tasks, such as code search and code summarization.
Furthermore, Guo et al.~\cite{DBLP:conf/iclr/GuoRLFT0ZDSFTDC21} proposed GraphCodeBERT to advance CodeBERT by incorporating data-flow information into pre-training.


\subsection{Code Clone Detection}
It is a fundamental task to detect similar code (or clone code) in many software engineering tasks (e.g, code reuse, code summarization, and bug detection). 
Code clones can be roughly categorized into the following types: Type-1, Type-2, Type-3 and Type-4.
CCFinder~\cite{kamiya2002ccfinder} extracted a series of tokens from code file and transform it according to several rules into a regular form for Type-1 and Type-2 clone detection.
NICAD~\cite{roy2008nicad} introduced a two-stage approach which first finds and regulates potential clones to remove noise using pretty-printing and then enumerate potential clones using dynamic clustering.
There are a variety of methods operation at different level to represent the syntax and semantic structure of a program. Jiang et al.~\cite{jiang2007deckard} proposed Deckard, which incorporates ASTs into code representation learning use locality sensitive hashing for efficient clustering. 
To detect Type-3 clone, SourcererCC~\cite{sajnani2016sourcerercc} was designed to capture the shared similarity of tokens among multiple approaches.

Recently, deep neural networks are applied to the task of code mathing.
For example, White et al.~\cite{white2016deep} proposed DLC, which takes the lexical and syntactic information of code into account, and designs Recurrent Neural Networks to represent them. 
To better express structured syntactic information of code,
Wei et al.~\cite{wei2017supervised} proposed to represent the syntactic information of code using TreeLSTM over ASTs.
Furthermore, Zhang et al.~\cite{zhang2019novel} decomposed the AST into sentence-based abstract syntax subtrees, and proposed a two-way loop network for representation. 
This method has achieved good results in code matching. 
Zhao et al.~\cite{zhao2018deepsim} proposed to consider the data flow and control flow of source code, and proposed a deep learning framework for code representation. 
Based on the control-flow graph of code, Wu et al.~\cite{wu2020scdetector} introduced a centrality analysis method from the perspective of social network analysis, which is efficient and effective in source code matching.
Zhang et al.~\cite{zhang2019novel} proposed ASTNN, an AST-based model for code embedding, which decomposes a big AST into a series of limited-scale statement trees. ASTNN has achieved promising performance in code clone detection.

\subsection{Cross-Language Source Code Analysis}
With the recent progress in transfer learning, transferring knowledge across different programming languages has become a promising research direction. 
Xia et al.~\cite{xia2014cross} studied the problem of cross-language bug localization based on language translation, which focuses on ranking source code files based on comments written in different natural languages. 
Chen et al.~\cite{DBLP:conf/nips/ChenLS18} proposed a tree-to-tree approach to transform programs from one language into another~\cite{DBLP:conf/nips/ChenLS18}.
Bui et al.~\cite{bui2019bilateral} proposed a bilateral model of two encoders, each of which is for encoding the abstract syntax of code in one programming language.
Bui et al.~\cite{DBLP:conf/sigsoft/BuiYJ19} proposed to improve program translation via mining API mappings across programming languages based on adversarial learning.
Nafi et al.~\cite{nafi2019clcdsa} proposed an approach for cross-language source clone matching based on structured syntactic features and code API documentation.
Gu et al.~\cite{DeepAM} proposed DeepAM, which can automatically mine API mappings between two languages from code corpus with single-language projects.
Our paper is the first to study binary-source code matching across different programming languages.

\section{Conclusion and Future Work\label{sec_conclusion}}
In this research, we have formulated a new problem of cross-language binary-source code matching. We also propose a novel approach, termed \tool, based on Transformer and intermediate representations of programs.
Comprehensive experiments are conducted on two tasks of cross-language binary-source code matching, and cross-language source-source code matching, over a created cross-language dataset.
Experimental results and analysis show that \tool significantly outperforms other state-of-the-art models. 
For the matching of Java binary code to C source code, when comparing with B2SFinder, \tool significantly improves the Recall, Precision, and F1 from 0.41, 0.35 and 0.38 to 0.55, 0.68 and 0.61, respectively.

Due to the challenging nature of the cross-language binary-source code matching problem, there is still ample room for enhancing the strength of our \tool. In our future work, we will design more effective mechanisms to further improve the accuracy of code matching.
We also plan to extend our approach to support smaller clone detection granularity, such as clone detection on a single function or small code snippet. 

\noindent\textbf{\textit{Artifacts.}} All the experiments in this paper are integrated into the open-source toolkit \textsc{NaturalCC}~\cite{wan2020naturalcc}.
Our datasets and source code used in this work are available at \textbf{\url{https://github.com/CGCL-codes/naturalcc}}.

\section*{Acknowledgements}

This work is supported by National Natural Science Foundation of China under grand No. 62102157, 61972444. This work is also partially sponsored by Tencent Rhino-Bird Focus Research Program of Basic Platform Technology.
We would like to thank all the anonymous reviewers for their constructive comments on improving this paper.

\bibliographystyle{IEEEtran}
\bibliography{ref}

\end{document}